\documentclass[11pt,twocolumn]{aastex6}
\usepackage{color}
\usepackage{epsfig}
\usepackage{setspace} 
\shorttitle{Distance to the LMC}
\shortauthors{Elgueta, S. S., Graczyk, D., et al.}
\begin{document}


\title{Orbital and physical parameters, and the distance of the eclipsing binary system OGLE-LMC-ECL-25658 in the Large Magellanic Cloud}
\author{S. S. Elgueta\altaffilmark{1,2}, D. Graczyk\altaffilmark{2,1,3}, W. Gieren\altaffilmark{1,2},  G. Pietrzy{\'n}ski\altaffilmark{3,1}, I. B. Thompson\altaffilmark{4},  P. Konorski\altaffilmark{5}, B. Pilecki\altaffilmark{3}, S.~Villanova\altaffilmark{1},
A. Udalski\altaffilmark{5}, I. Soszy{\'n}ski\altaffilmark{5}, K. Suchomska\altaffilmark{5}, P. Karczmarek\altaffilmark{5}, M. G{\'o}rski\altaffilmark{2,1}, P. Wielg{\'o}rski\altaffilmark{3}}
\affil{$^1$Universidad de Concepci{\'o}n, Departamento de Astronomia, Casilla 160-C, Concepci{\'o}n, Chile}
\affil{$^2$ Millennium Institute of Astrophysics (MAS), Chile}
\affil{$^3$ Nicolaus Copernicus Astronomical Center, Polish Academy of Sciences, ul. Bartycka 18, 00-716 Warszawa, Poland}
\affil{$^4$ Carnegie Observatories, 813 Santa Barbara Street, Pasadena, CA 911101-1292, USA}
\affil{$^5$ Warsaw University Observatory, Al. Ujazdowskie 4, PL-00-478, Warszawa, Poland}

\begin{abstract}

We present an analysis of a new detached eclipsing binary, OGLE-LMC-ECL-25658, in the Large Magellanic Cloud. 
The system consists of two late G-type giant stars on an eccentric orbit and orbital period of $\sim 200$ days. 
The system shows total eclipses and the components have similar temperatures, making it ideal for a precise distance determination. 
Using multi-color photometric and high resolution spectroscopic data, we have performed an analysis of light  
and radial velocity curves simultaneously using the Wilson Devinney code. We derived orbital and physical parameters 
of the binary with a high precision of $<1$ \%. The masses and surface metallicities of the components are virtually the same 
and equal to $2.23\pm0.02\, M_\odot$ and [Fe$/$H$]=-0.63 \pm 0.10$ dex. However their radii and rates of rotation show a
distinct trace of differential stellar evolution. The distance to the system was calculated using an infrared calibration 
between V-band surface brightness and $(V\!-\!K)$ color, leading to a distance modulus of
$(m\!-\!M) = 18.452 \pm 0.023$ (statistical) $\pm 0.046$ (systematic). Because OGLE-LMC-ECL-25658 is located relatively far 
from the LMC barycenter we applied a geometrical correction for its position in the LMC disc using the van der Marel et al. model of the LMC. 
The resulting barycenter distance 
to the galaxy is $d_{\rm LMC}=50.30\pm0.53$ (stat.) kpc, and is in perfect agreement with the earlier result of Pietrzy\'nski et al.~(2013). 
\end{abstract}

\keywords{Stars: Eclipsing Binaries, Galaxies: distances, LMC}

\section{Introduction}
The distance to the Large Magellanic Cloud (LMC) is of great astrophysical importance for at least two reasons.
Firstly, the LMC serves as the best anchor to establish the zero point of the extragalactic distance scale, and a variety
of distance indicators, including classical Cepheids, RR Lyrae stars, red clump stars, and the Tip of the Red Giant Branch
(TRGB) method can be calibrated in the LMC with better accuracy than in any other galaxy, at the present time. Secondly,
the LMC represents an ideal laboratory for the detailed study of different stellar populations, given its distance of
about 50 kpc which is far enough to see, in a first approximation, all its objects at the same distance, and at the
same time close enough to study in great detail even its faint stellar populations. Given these advantages, many
astrophysical studies of LMC stars, stellar clusters and gaseous regions have been carried out over the past century;
in particular, many attempts have been made to determine a precise and accurate distance to the LMC in order to put
the luminosities and sizes of its components onto an absolute scale. It has proved extremely difficult to establish
an accurate and reliable distance to the LMC, given the different, and often large systematic uncertainties on the
different methods which have been applied to solve this task.

A breakthrough came with the discovery of a very special and rare class of eclipsing binaries in the Magellanic Clouds
by the OGLE project \citep{ud98,wyr03,wyr04}. These systems consist of two usually quite similar red giant components, 
and offer two very important advantages over the (usually brighter and thus easier-to-observe) early-type eclipsing binaries 
in the context of distance determination: they can be easily and precisely
analyzed using standard techniques and without resorting to uncertain theoretical predictions, and the angular diameters
of the component stars, and hence their distances, can be precisely determined from a surface brightness - color relation
 which is accurately determined for late-type stars \citep[e.g.][]{ker04,dib05}, but not for
early-type stars \citep[e.g.][]{cha14}. A first system of this class has been analyzed by \cite{pie09},
and led to a distance accurate to 3\%, including the (well-determined) systematic uncertainties. Due to the hard work 
of the Araucaria Project team, four years later data for these
hard-to-observe systems (due to their long orbital periods and relative faintnesses) became ready for the analysis of eight
late-type systems; their analysis led to a LMC barycenter distance being accurate to 2.2\% \citep[][hereafter P13]{pie13} 
which currently represents the most accurate distance to the LMC ever measured.

In an effort to improve the LMC distance determination even further, mostly in the quest of improving the Hubble constant
to an accuracy of 1\% with the Cepheid-SN Ia method \citep[e.g.][]{rie11}, our group is currently observing more late-type systems
in the LMC. One of these systems, OGLE-LMC-ECL-25658, is particularly useful for deriving a precise distance: it is relatively
bright, both components have very similar temperatures, and the deep and total eclipses allow for very precise determinations
of the stellar radii and dynamical parameters. The ($V\!-\!K$) colors of the component stars allow, in turn, for a very precise determination
of their surface brightnesses, and in tandem with the radii their distances. Some basic information on the system
(coordinates, observed magnitudes and the orbital period) is given in Table 1. The system is different from those systems
studied in P13 in the sense that its location in the LMC is quite far (3.5 degrees) from the
geometrical center of the LMC and the line of nodes \citep[the model of][]{vMa02}. We present a detailed
analysis of this system in this paper, derive its orbital parameters and very accurate physical parameters for the
component red giants in the system, and derive its distance which, corrected with the geometrical model of van der Marel et al., strengthens
the LMC distance determination with late-type eclipsing binaries of P13. The results also add to our database of physical
parameters of red giants (masses to better than 1\%, radii, luminosities) in the Magellanic Clouds, which eventually will allow a much
improved understanding of the physics, and stellar evolution of red giants in environments of different metallicities.

The paper is organized in the following way: in section~2, we present and discuss the different sets of observations we used
for this study. In section~3 we describe the details of data analysis and in section~4 we give a summary of derived physical parameters. 
Section~5 describes the distance determination, and section~6 contains the conclusions and final remarks.

\section{Observations}
\subsection{Photometry}
\begin{figure}
\centering
\includegraphics[width=0.47\textwidth]{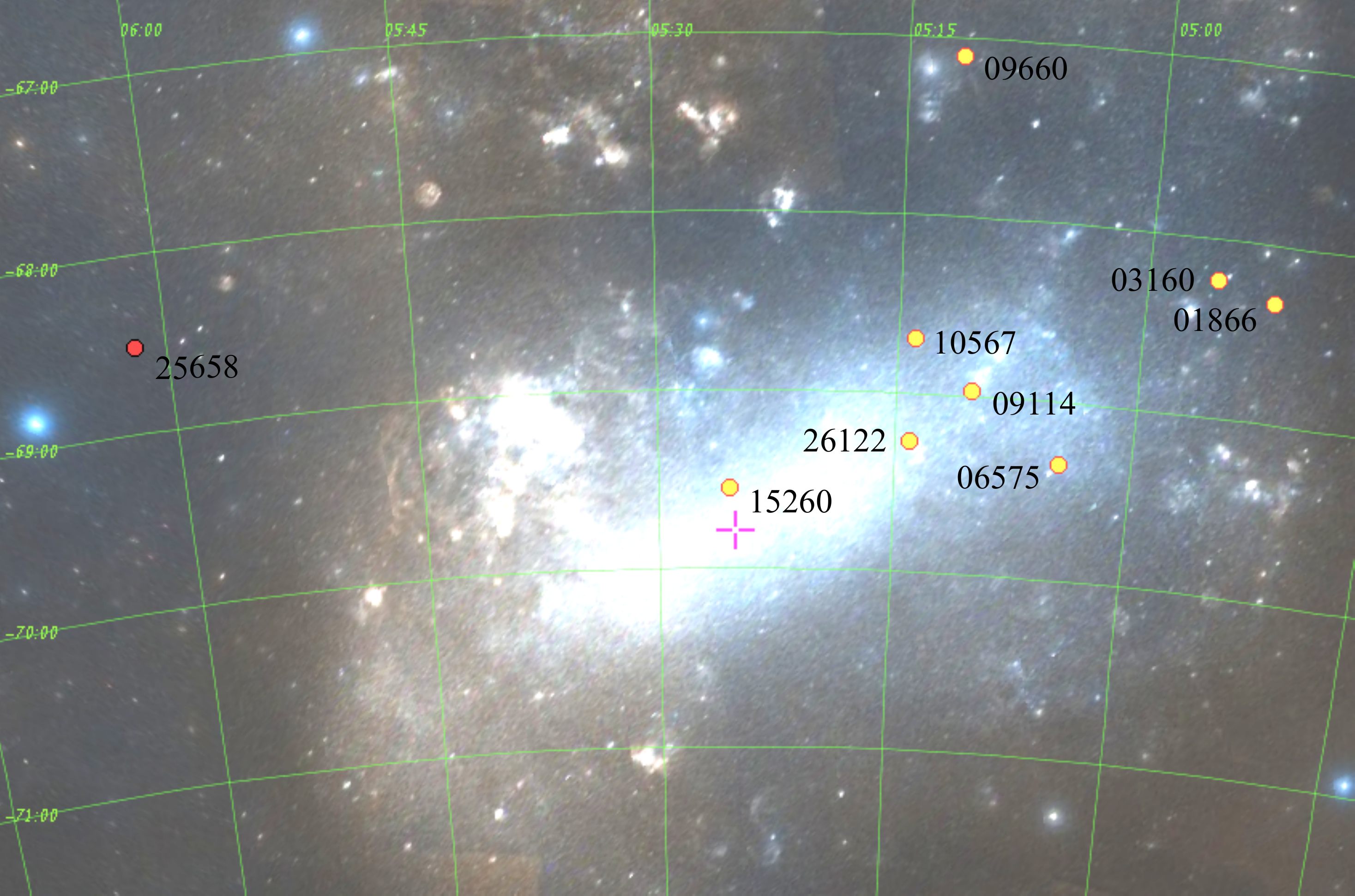}
\caption{Position of OGLE-LMC-ECL-25658 system in the LMC. It is located to the east part of galaxy 3.5 degree from its geometrical center (cross). Positions of the systems from P13 are also denoted. Image and positions of the systems obtained from Aladin Sky Atlas \citep{bon00}. North is up and East is to the left. \label{fig1} }
\end{figure}

The system was detected during the course of the OGLE III project and identified as an eclipsing binary system by \cite{gra11}. 
Its position relative to the LMC is shown in Figure~\ref{fig1}. Optical photometry in the Johnson-Cousins filters was obtained with 
the Warsaw 1.3 m telescope at Las Campanas Observatory in the course of the third \citep{ud03} and fourth \citep{sos12,ud15} phase of the Optical Gravitational
Lensing Experiment (OGLE) project \citep{ud97}. The raw data were reduced with the image subtraction technique
\citep{wo00,ud03} and instrumental magnitudes were calibrated onto the standard system using Landolt standards. The zero point error
of our optical photometry is 0.010 mag. The I band light curve shows a small amount of intrinsic variability, visible as light oscillations
with an amplitude of up to 0.01 mag, and a quasi-period of 50-100 days.

\begin{deluxetable}{@{}llc@{}} 
\tablecaption{Basic Data for  OGLE-LMC-ECL-25658  \label{tbl:1}}
\tablehead{
 \colhead{} & & \colhead{Reference}}
\startdata
OGLE III (internal)		&   LMC204.6  7868  & 1 \\
MACHO				&   75.13255.26	& 2 \\
EROS-2				&   lm0254m23176    & 3 \\
\\
R.A. (J2000)  &  $\;\;\;$06:01:58.77 & 1 \\
DEC (J2000) &  $-$68:30:55.1& 1 \\  
V  (mag) & 16.995 $\pm$ 0.012 & 4 \\
I (mag)  & 15.870 $\pm$ 0.010 & 4 \\
J$_{J}$ (mag)&15.071 $\pm$ 0.023& 4\\
H$_{\rm 2MASS}$ (mag) &14.447 $\pm$ 0.055& 5\\
K$_{J}$ (mag)&14.357 $\pm$ 0.019& 4\\
Orbital period (days) & 192.789 $\pm$ 0.001 & 4 
\enddata
\tablecomments{References: 1 - \cite{gra11}, 2 - \cite{fra08}, 3 - \cite{kim14}, 4 - this work, 5 - \cite{skr06}}
\end{deluxetable}   

Near-infrared photometry was collected with the ESO NTT telescope on La Silla, equipped with the SOFI camera. In total we obtained 22 epochs of infrared photometry for our system outside eclipses. Our photometry was tied to the UKIRT system by observations of a number of JHK standards from \cite{haw01} and then transformed onto the Johnson system using the equations given by \cite{car01} and \cite{BB98}. The zero point photometric error of our IR photometry is 0.015 mag. 

\subsection{Spectroscopy}
High resolution echelle spectra were collected with the Clay 6.5 m telescope at Las Campanas equipped with the MIKE spectrograph, and with the 3.6 m telescope at ESO-La Silla, equipped with the HARPS spectrograph. Details of the setup are given in \cite{gra14}. In total we collected 9 HARPS spectra and 15 MIKE spectra. In some same cases (e.g. weak S/N) we could use only one part of a MIKE spectrum (blue or red). The HARPS spectra were reduced and calibrated with the standard on-site pipeline whereas the MIKE spectra were reduced with the pipeline software developed by D. Kelson, following \cite{kel03}. In order to determine radial velocities of the components, we employed the Broadening Function (BF) formalism introduced by \cite{ru92,ru99}. Radial velocities were derived using the RaveSpan software \citep{pi12}. Absorption line profiles of the cooler and larger component (the secondary) are significantly rotationally broadened while the line profiles of the primary are almost gaussian. We determined total line broadenings in term of rotational velocity of the components to be $v_1\sin{i}=9.5\pm1.0$ km s$^{-1}$ and  $v_2\sin{i}=18.9\pm1.2$ km s$^{-1}$.
During the preliminary fitting with the RaveSpan software we detected a systematic shift between the HARPS and MIKE radial velocities amounting to +0.76 km s$^{-1}$. This shift was subtracted from all radial velocities measured from MIKE spectra. Table~\ref{rad:vel} presents all our radial velocity measurements used in the subsequent analysis.   


\begin{deluxetable}{@{}cccl@{}} 
\tablecaption{Radial velocities of OGLE-LMC-ECL-25658  \label{rad:vel}}
\tablehead{
 \colhead{HJD} & \colhead{RV1} & \colhead{RV2} & \colhead{Spectrograph}}
\startdata
2454784.82235 & 288.315 & 226.897 & MIKE \\
2454809.75120 & 279.247 & 235.338 & HARPS \\
2454816.71919 & 275.228 & 238.456 & MIKE \\
2454883.63273 & 239.947 & 274.251 & MIKE \\
2454887.73301 & 238.783 & 276.555 & HARPS \\
2454888.71727 & 237.623 & 276.534 & HARPS \\
2454889.72089 & 237.440 & 277.668 & HARPS \\
2455088.89846 & 234.038 & 281.724 & HARPS \\
2455185.82555 & 283.472 & 230.683 & HARPS \\
2455218.85162 & 267.462 & 247.478 & MIKE-Blue \\
2455219.85556 & 267.605 & 248.489 & MIKE-Blue \\
2455272.74222 & 238.997 & 275.949 & MIKE-Red \\
2455449.89368 & 247.642 & 267.928 & HARPS \\
2455470.80855 & 235.760 & 279.769 & HARPS \\
2455502.73785 & 223.229 & 291.191 & HARPS \\
2455557.60017 & 287.778 & 226.300 & MIKE \\
2455590.70444 & 274.517 & 240.392 & MIKE \\
2455591.70406 & 274.386 & 241.211 & MIKE \\
2455882.83047 & 223.689 & 291.153 & MIKE \\
2455883.74846 & 222.711 & 290.662 & MIKE \\
2455950.75924 & 285.780 & 229.107 & MIKE \\
2455952.74407 & 285.202 & 229.428 & MIKE \\
2455964.78753 & 280.053 & 235.016 & MIKE-Red
\enddata
\tablecomments{Mean error of radial velocity determination is 300 m s$^{-1}$
for HARPS and 350 m s$^{-1}$ for MIKE.}
\end{deluxetable}   

\section{Modeling Strategy}
In order to derive absolute fundamental parameters for the system we used the Wilson-Devinney (hereafter WD) code version 2007 \citep{WD71,wil79,wil90,wil07} which allows us to analyze multi-band light curves and radial velocity curves simultaneously. This code also implements the automated differential correction (DC) optimizing subroutine helping us to get an optimal model of our binary system. As the system is well detached we utilized the JKTEBOP program \citep{sou04a,sou04b}, based on the EBOP code \citep{pop81}, to get a reliable determination of the statistical errors by Monte Carlo simulations. 

We follow the metodology from \cite{gra14}. The first step of the analysis uses the information obtained from the combined light and radial velocity curves plus a first reddening estimate to estimate basic physical parameters (e.g. surface temperature) and subsequently decompose the components' spectra. The second step utilizes information obtained from an atmospheric analysis of the decomposed spectra in order to obtain a full set of internally consistent parameteres of the system.

\subsection{Initial parameters}
The detached configuration (Mode 2) was chosen during the whole analysis and a simple reflection treatment (MREF = 1, NREF = 1) was employed. A stellar atmosphere formulation was selected for both stars (IFAT = 1). Level dependent weighting was applied (NOISE = 1), and curve dependent weightings (SIGMA) were calculated after each iteration. A logarithmic limb darkening law was used \cite{kli70} with coefficients automatically computed from the \cite{vh93} tables (option LD= $-$2). The orbital period $P$ and epoch of the primary minimum were taken from the OGLE-III catalogue of eclipsing binaries in the LMC \citep{gra11}. Initially we assumed a temperature of $T_1=5000$ K for the primary component (i.e.~component eclipsed during primary minimum). The albedos $A_{1,2}$ and the gravity brightenings $g_{1,2}$ were set to standard values expected for a cool convective atmosphere. The rotation parameter of the two components was set to $F_{1,2}=1$, i.e. synchronous rotation. The orbit is significanlty eccentric and using the LC module of the WD code we computed a number of trial light curves and radial velocity curves to obtain good starting values of stellar radii and the orbital parameters. As the eclipses are total, and the separation of the components is relatively large the orbital inclination $i$ must be close to 90 degrees.     

\subsection{Initial fitting}
\label{init:fit}
We fitted simultanously two optical light curves (V and I bands) and two radial velocity curves corresponding to each of the components. After each iteration curve dependent weights were updated. As free parameters of the WD model we chose the semimajor axis $a$, the orbital eccentricity $e$, the argument of periastron $\omega$, the phase shift $\phi$, the systemic radial velocity $\gamma$, the orbital inclination $i$, the secondary star average surface temperature $T_2$, the modified Roche lobe surface potential of both components $\Omega_{1,2}$, the mass ratio $q = M_2/M_1$, the observed orbital period $P_{obs}$, and the relative monochromatic luminosity of the primary star in the two bands $L1_V$, $L1_I$. It is worth noticing that within the WD code there is no possibility to directly adjust the radial velocity semiamplitudes K1 and K2; instead, the semimajor axis and the mass ratio are adjusted simultaneously. For all fitted parameters we use a parameter increament, required by the DC procedure, equal to 0.7\% of the nominal value. 

From the initial fitting it was clear that the orbital inclination is so close to 90 degrees that it causes numerical instability and some problems with a solution convergence. We had to repeat the fitting with different increments of $i$ until we could find a satisfactory solution. The temperature of the secondary star $T_2$ was found to be very similar to $T_1$. We updated the temperature estimate of the primary $T_1$ as follows. We determined the temperature ratio of the components to be equal to $T_2/T_1= 0.972$. The extinction in the direction of our system was estimated from reddening maps of the Magellanic Clouds \citep{has11}. We calculated $E(B\!-\!V)=0.091\pm 0.030$ mag using the equation:

\begin{equation}
E(B-V)=\frac{E(V-I)}{1.3} + 0.057 
\end{equation}
where $E(V\!-\!I)$ is the color excess from the reddening map, denominator of 1.3 is adopted from \cite{bes98} and $\Delta(B\!-\!V) = 0.057$ mag is the foreground galactic reddening in the direction of the system as derived from the dust maps of \cite{sch11}. Then we corrected the V,I,K band infrared magnitues of the system for exctinction using the interstellar extinction law given by \cite{car89} and \cite{odo94} assuming $R_{\rm V}=3.1$. Using a number of calibrations between $(V\!-\!I)$, $(V\!-\!K)$ colors and temperature \citep{dib98,alo99,hou00,ra05,gon09,cas10,wor11} and temperature ratio, after a few iterations we determined $T_1=4825$ K. A model of the system with this adopted temperature was subsequently utilized in obtaining the components' decomposed spectra.

\subsection{Spectral disentangling process }
Each of our recorded spectra are composite spectra consisting of the spectral features of both components. A time series analysis of this composite spectrum can give us information about the dependence of the radial velocities of each component on the orbital phase, and spectral features of the component stars. This meaningful information is "entangled", and requires a specific process for its extraction. In order to constrain the atmospheric parameters of our binary components, especially their surface temperatures, a spectral disentangling process was performed. We followed the method described by \cite{gon06} to disentangle individual spectra of the binary components. The method works in the real wavelength domain. It requires rather high signal to noise ratio spectra to work properly, and thus, we were restricted to using only the red part of MIKE spectra. We used the two-step method described in detail by \cite{gra14} to derive renormalized disentangled spectra. Renormalization was done using optical light ratios calculated with the preliminary model of the system (see Section \ref{init:fit}). The resulting spectra have S$/$N$\sim50$ at 5500 ${\rm \AA}$.

\subsection{Atmospheric analysis}
The decomposed spectra were used for deriving the basic atmospheric parameters like effective temperatures $T_{\rm eff}$, microturbulence $\nu_{\rm t}$ and metallicity [Fe$/$H].  By measuring the equivalent widths (EWs) of the iron spectral lines it is possible to obtain the iron content of the components \citep{mar08,vil10}.  Surface gravities $\log{g}$ of both stars were kept at values corresponding to radii and masses derived from preliminary the WD solution. The solar iron abundance chosen was $log \epsilon (Fe) = 7.50$. Atmospheric parameters were obtained as it follows: model atmospheres were calculated using ATLAS9 \citep{kur70} assuming as initial estimations for $T_{\rm eff}$ and $\nu_{t}$ values typical for giant stars (4800 K and 1.80 km s$^{-1}$). Lines of Fe I and Fe II were used for this purpose. The [Fe$/$H] value of the model was changed at each iteration according to the output of the abundance analysis. The typical accuracy of the parameters are 70 K, 0.10 dex, and 0.2 km s$^{ - 1}$ for $T_{\rm eff}$, [Fe$/$H] and $\nu_{t}$, respectively. The resulting atmospheric parameters for our system are summarized in Table ~\ref{tbl:2}. 

\begin{table}
\begin{center}
\caption{Atmospheric Parameters } \label{tbl:2}
\begin{tabular}{l  c  c  c} 
\hline
    & $T_{\rm eff}$  & [Fe$/$H] & $\nu_{t}$ \\
\hline
Primary  & 4860 & $-$0.65 & 1.70\\
Secondary & 4730 & $-$0.62 & 1.80 \\
\hline
\end{tabular}
\end{center}
\end{table}    

\subsection{Final solution}
For the fine-tuning of the model we set the temperature of the primary to the value derived from the atmospheric analysis, $T_{1}=4860$ K, and we then recalculated the model. The resulting optical light ratios were in full agreement with light ratios from the preliminary solution and we do not iterate the renormalization of decomposed spectra and the atmospheric analysis. The temperature scale consistency of the system was checked by computing the distance to the system resulting from a scaling of the bolometric flux observed at Earth. To calculate the bolometric corrections, we used an average from several calibrations \citep{flo96,alo99,cas10}. These distances were then compared to the distance computed with a surface brightness - color relation (see Sect.~\ref{sec:dist}). The good agreement between those two distances put confidence into our adopted temperature scale. 

\begin{figure}
   \centering
   \includegraphics[width=\hsize]{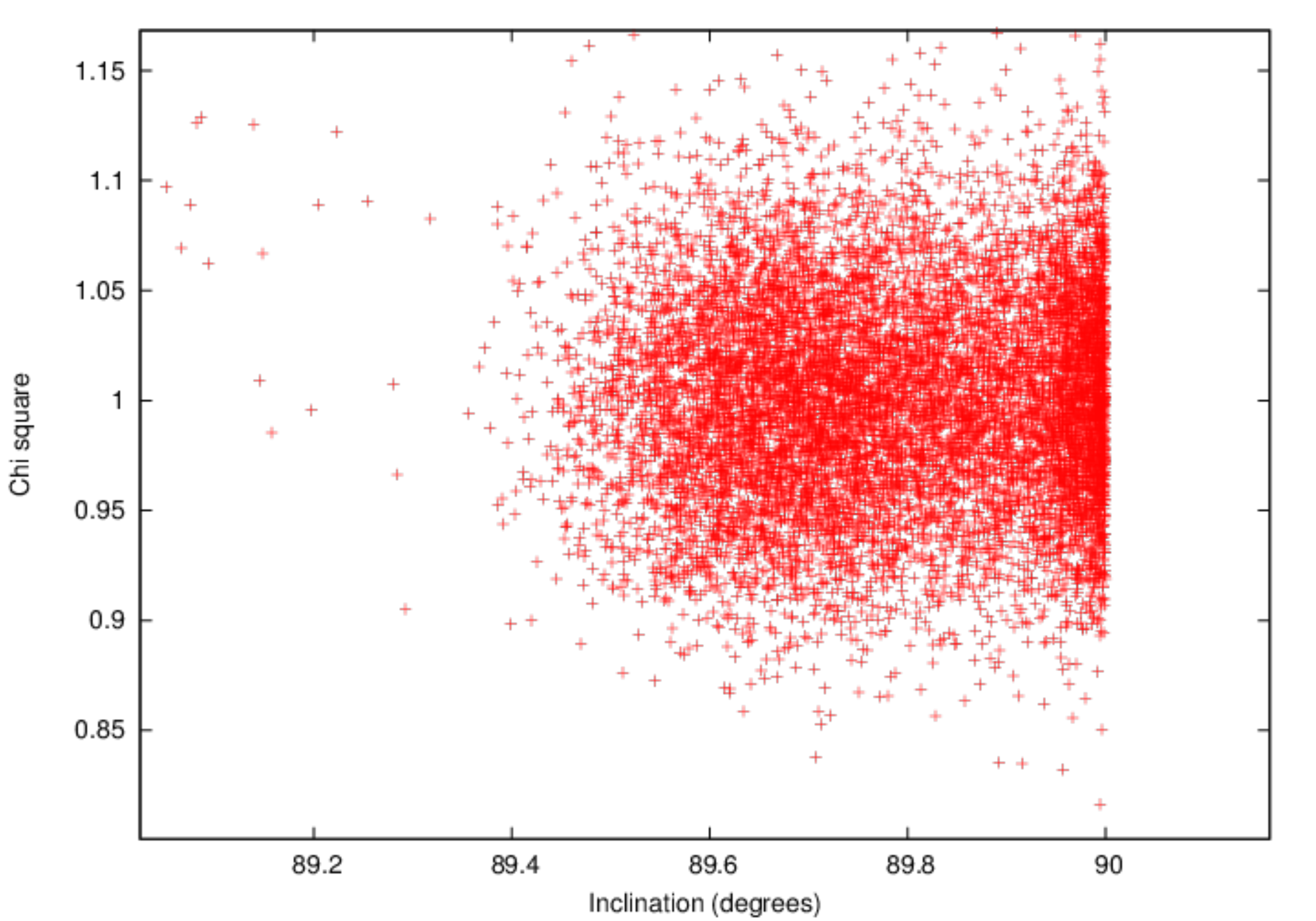}
      \caption{$\chi^2$ map of the orbital inclination. \label{fig6}}
         \label{chiinc}
   \end{figure}

\begin{figure*}
\centering
\includegraphics[scale=0.54]{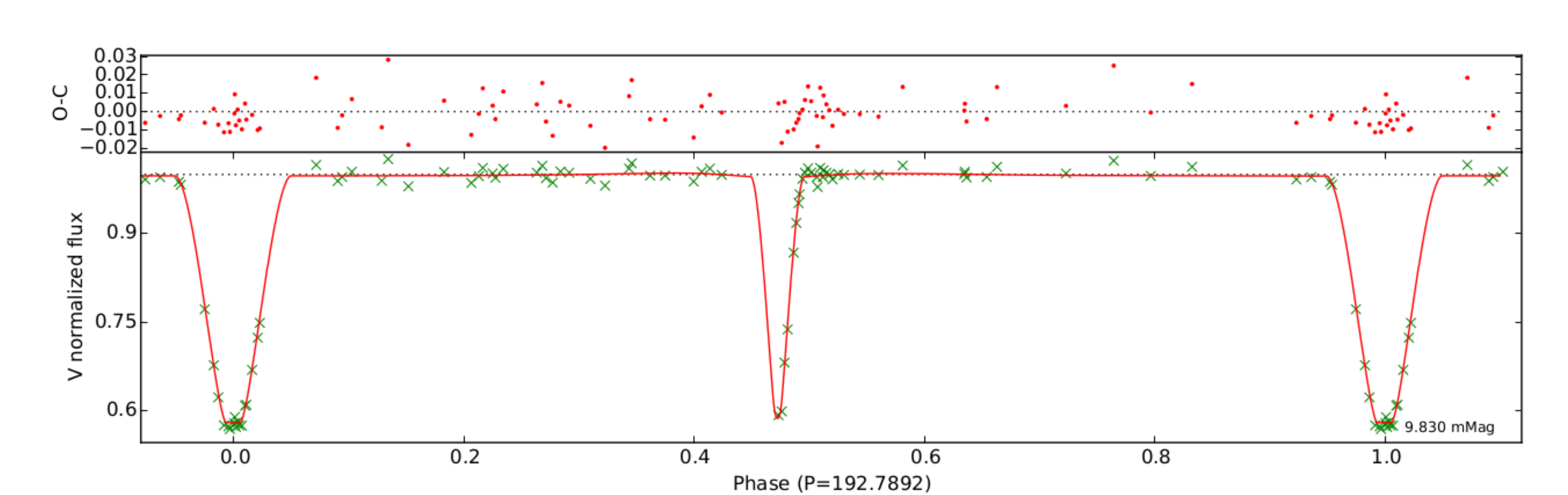} 
\includegraphics[scale=0.54]{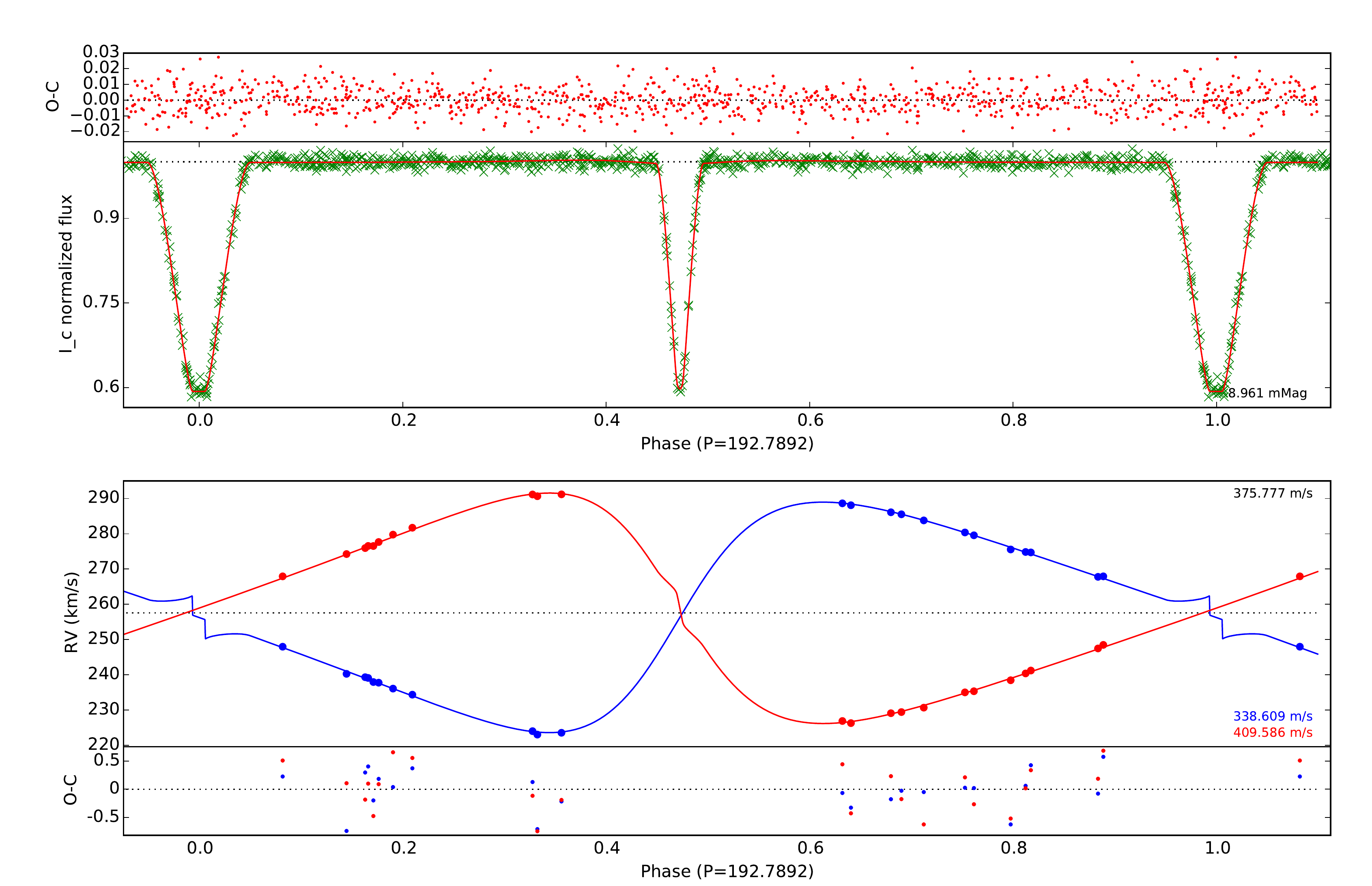}
\caption{The simultaneous WD solution for $V$-band light curve ({\it upper panel}),  $I_C$-band light curve ({\it middle panel}) and the radial velocity curves ({\it lower panel}). \label{fig2}}
\end{figure*}

The final tuning of the model was impeded by notorious numerical instability of the WD code close to the 90 degree limit of the orbital inclination. In order to get a definitive value for the inclination and also to find reliable statistical errors on the photometric parameters we ran extensive Monte Carlo simulations using the JKTEBOP program \citep{sou04a,sou04b}, based on EBOP code \citep{pop81}. The system is well detached and biaxial representation of stellar surface utilized in EBOP is sufficiently precise to not introduce systematics in the solution. The task 8 was choosen and we calculated 10000 models based on the I band light curve. Figure~\ref{fig6} shows a $\chi^2$ map for the orbital inclination and illustrates the indeterminecy of this parameter. If fact all inclinations larger than 89.6 deg are allowed. The best solution returned from the MC simulations gives $i=89.79$ deg and this value was adopted in our final WD runs. The solutions from the WD and JKTEBOP codes are very consistent and show only small, insignificant differences. Table~\ref{tab:par} contains basic parameters of our final solutions with the WD code. Resulting synthetic light and radial velocity curves and a comparison with the data is presented in Fig.~\ref{fig2}.  


Also, we computed two additional sets of models by adjusting: (1) the four coefficients of the linear law of limb darkening (mode LD = +1) for both stars and both light curves, (2) the third light $l_3$ in V and I band. We checked the resulting reduced $\chi^2$ but we didn't see any improvements. Moreover, the obtained third light values were consistent with zero, leading us to adopt the configuration with $l_3=0$ and with tabulated limb darkening coefficients for logarithmic law. 


\begin{table}
\begin{center}
\caption{Solution parameters from the WD code\label{tab:par}}
\begin{tabular}{@{}lr@{$\,\pm\,$}l@{}}
\hline
\multicolumn{3}{c}{Orbital Parameters} \\
\hline
Orbital period $P_{\rm obs}$ (d) & 192.7892 & 0.0014 \\
$T_{0}$  (HJD - 2450000) &  3891.605 & 0.009\tablenotemark{a} \\
$a \sin{i} $ ($R_{\sun}$) & 231.237 & 0.610 \\
Systemic velocity $\gamma_1$ (km s$^{-1}$)  & 257.25 & 0.06  \\
Systemic velocity $\gamma_2$ (km s$^{-1}$)  & 257.58 & 0.06 \\
Periastron longitude $\omega$(deg)  & 263.93 & 0.22  \\
Eccentricity $e$  & 0.3731 &  0.0046 \\
Mass ratio $M_{2}/M_{1}$   & 1.0001 & 0.0048\\
\hline
\multicolumn{3}{c}{Photometric Parameters} \\
\hline
Orbital inclination $i$ (deg) &  89.79 & 0.18\tablenotemark{b}\\
Surface potential $\Omega_1$ &12.394 & 0.040 \\
Surface potential $\Omega_2$ &9.985 & 0.050 \\
Temperature ratio $T_{2}/T_{1}$  &  0.9714 & 0.0012   \\
Relative radius $r_{1}$   &  0.0927  & 0.0006\\
Relative radius $r_{2}$  & 0.1193 &  0.0010 \\
Light ratio $(L_{2}/L_{1})_{V}$ &  1.3953 & 0.0050  \\
Light ratio $(L_{2}/L_{1})_{I}$ &   1.4681 & 0.0046  \\
Light ratio $(L_{2}/L_{1})_{J}$ & \multicolumn{2}{c}{1.5365\tablenotemark{c}} \\
Light ratio $(L_{2}/L_{1})_{K}$ & \multicolumn{2}{c}{1.5964\tablenotemark{c}} \\
 \hline
 \multicolumn{3}{c}{Derived Quantities} \\
\hline
Rest frame $P_{\rm orb} $   &   192.6237 & 0.0014  \\
Semimajor axis $a$ ($R_{\sun}$)&  231.040 & 0.612\tablenotemark{d}  \\
Velocity semiaplitude $K_{1}$ (km s$^{-1}$)  & 32.70 & 0.13 \\
Velocity semiaplitude $K_{2}$ (km s$^{-1}$)  & 32.70& 0.13 \\ 
k = $r_{2}/r_{1}$  &  1.287 & 0.014 \\
$r_{1} + r_{2}$  & 0.2120 & 0.0011  \\
\hline
\end{tabular}
\end{center}
$^{a}$ {\small Epoch of the primary minimum}\\
$^{b}$ {\small From the MC calculations with JKTEBOP} \\
$^{c}$ {\small Extrapolated from the WD code} \\
$^{d}$ {\small Calculated from rest frame orbital period}
\end{table}

\section{Absolute dimensions}
A summary of the derived physical parameters of our system is given in Table~\ref{tab:abs}.
Spectral types of the components were estimated from temperatures using calibration by \cite{alo99}.
The observed orbital period of the system $P_{\rm obs}$ and the true orbital period $P_{\rm orb}$ are linked through a relation:
\begin{equation}
P_{\rm obs}=P_{\rm orb}(1+ \frac{\gamma}{c})
\end{equation}
where $c$ is the speed of light. The corrected period $P_{\rm orb}$ is used to calculate the semimajor axis of the system.  The masses were derived from the equations: 
\begin{equation}
M_{1}[M_\sun]= 1.34157 \times 10^{-2} \frac{1}{1+q} \frac{a^{3} [R_{\sun}]}{P^{2} [\rm d]} 
\end{equation}
\begin{equation}
M_{2}[M_{\sun}] = M_{1} \cdot q   
\end{equation}

Projected rotational velocities $v \sin{i}$ of the components were calculated from the observed line broadening taking into account the effects of the macroturbulence $\nu_M$ and the instrumental broadening $\nu_{\rm ib}$. To estimate the macroturbulence we used the calibration given by \cite{mas08}, and the instrumental profile broadening for MIKE's resolving power of R=40000 was estimated to be 4.5 km s$^{-1}$. Comparison of the resulting $v \sin{i}$ values with synchronous and pseudo-synchronous rotational velocities shows that the rotation of the smaller, primary component seems to be locked to the orbital period (synchronism) while the secondary rotates much faster in accordance with the periastron rotation lock. In such a comparison we assume that both stellar axes of rotation are perpendicular to the orbital plane. However, this mismatch of rotational velocities may also mean a strong misalignment of the rotation axes, with the primary's axis being tilted against the orbital plane by $\sim30$~deg. 




\begin{table}
\begin{center}
\caption{Physical Properties of OGLE-LMC-ECL-25658 \label{tab:abs}}
\begin{tabular} {@{}lr@{$\,\pm\,$}lr@{$\,\pm\,$}l@{}}
\hline
Property & \multicolumn{2}{c}{Primary} & \multicolumn{2}{c}{Secondary} \\ \hline
Spectral Type &   \multicolumn{2}{c}{G8 III}  &  \multicolumn{2}{c}{G9 III}   \\ 
Mass $M$ ($M_{\sun}$) &  2.229 & 0.019  &  2.230 & 0.019\\ 
Radius $R$ ($R_{\sun}$) &  21.41 & 0.15  &  27.57 & 0.24\\ 
Gravity $\log{g}$ (cgs)  &  2.125 & 0.006  & 1.906 & 0.007 \\ 
$v \sin{i}$ (km s$^{-1}$) & 6.7 & 1.4 & 17.7 & 1.3 \\
$v_{\rm sync}$ (km s$^{-1}$) & 5.62&0.04 & 7.24 & 0.06\\
$v_{\rm psync}$ (km s$^{-1}$) &13.3 & 0.2& 17.1 & 0.3\\
Temperature $T_{\rm eff}$ (K) & 4860 & 70 & 4721 &  75 \\ 
Luminosity\tablenotemark{a} ($L_{\sun}$ ) &  230 & 14 & 339 &  22 \\ 
$M_{\rm bol}$\tablenotemark{b} (mag) &  $-$1.15 & 0.06   & $-$1.58 & 0.07 \\ 
$M_{V}$ (mag) &  $-$0.81 & 0.07 & $-$1.17&  0.08\\ 
$[$Fe$/$H$]$ (dex) & $-$0.65 & 0.10 & $-$0.62  & 0.10 \\ 
$E(B\!-\!V)$ (mag)&   \multicolumn{4}{c}{$0.091 \pm 0.030$}  \\ 
$(m\!-\!M)$\tablenotemark{c} (mag) & \multicolumn{2}{c}{18.452}& \multicolumn{2}{c}{18.453}\\
Distance\tablenotemark{d}  (kpc) & \multicolumn{4}{c}{$49.03 \pm 0.53$(stat.)$\pm1.04$(syst.)} \\
Distance to the LMC (kpc) & \multicolumn{4}{c}{$50.30 \pm 0.53$ (stat.)} \\\hline
\end{tabular}
\end{center}
$^{a}$ {\small Assuming $T_\odot=5777$ K}\\
$^{b}$ {\small Assuming Sun's bolometric magnitude $+4.75$ mag}\\ 
$^{c}$ {\small The distance modulus}\\
$^{d}$ {\small Taking into account systematic uncertainty in the distance modulus of 0.046 mag}
\end{table}

\section{Distance}
\label{sec:dist}
For late-type stars we can use the very accurately calibrated (2 \%) relation between their surface brightness and ($V\!-\!K$) color \citep{dib05} to determine their angular sizes from optical ($V$) and near-infrared ($K$) photometry. From this surface brightness-color relation (SBCR) we can derive angular sizes of the components of our binary systems directly from the definition of the surface brightness. Therefore the distance can be measured by combining the angular diameters of the binary components derived in this way with their corresponding linear dimensions obtained from the analysis of the spectroscopic and photometric data. The angular diameter of a star predicted by the surface-brightness color relation is:

\begin{equation}
\theta = 10^{0.2 (S - m_{0} )} 
\end{equation}
where S is the surface brightness in a given band and $m_{0}$ is the unreddened magnitude of a given star in this band. The distance in parsecs then follows directly from angular diameter scaling and is given by a simple linear equation:

\begin{equation}
d [{\rm pc}] = 9.300 \frac{R [R_{\sun}]} { \theta[{\rm mas}]}.
\end{equation}

The individual, extinction corrected magnitudes of the components were calculated using the light ratios from Table~\ref{tab:par}. To estimate the angular diameters we utilized the SBCR calibration given by \cite{dib05}. The distance to the OGLE-LMC-ECL-25658 system was assumed to be an average of the distances determined to both components. Its value is 49.03 kpc and it corresponds to a true distance modulus of $(m\!-\!M)=18.452$. The difference between the components' individual distance moduli is smaller than 0.001 mag. Because the system is located significantly away from the center of the LMC, it is neccesary to calculate a geometric correction to find the distance to the barycenter of the LMC. To this end, we adopted the model of the LMC disc from \cite{vMa02} i.e. inclination of the disc to the plane of the sky $i=28^{\circ}$ and the position angle of the line of nodes $\Omega=128^{\circ}$. The correction corresponding to the position of the system is $\Delta d=+1.27$ kpc, translating into a barycenter distance to the LMC of $d_{\rm LMC}=50.30$ kpc, or a true distance modulus of $(m\!-\!M)_{\rm LMC}=18.508\pm0.023$ mag (statistical error). This value is in excellent agreement with the distance modulus to the LMC derived by P13 who reported $(m\!-\!M)=18.493\pm0.008({\rm stat.})\pm0.047({\rm syst.})$ based on eight late-type eclipsing binaries.   

\subsection{Error budget}
Contributions to the statistical error on the distance modulus determination are: uncertainty on the sum of the radii (0.012 mag), uncertainty on exctinction (0.013 mag), error of semimajor axis (0.008 mag), out-of-eclipse variability (0.010 mag), and combined error on photometry (0.006 mag). Combination of these contributions in quadrature yield 0.023 mag. We assumed this value as the total statistical uncertainty on the distance.

The sources of systematic uncertainty in our method are the following: the uncertainty on the empirical calibration of the surface brightness-color relation of di Benedetto (2005) (0.041 mag), uncertainty on the extinction (0.010 mag), metallicity dependence on the surface brightness-color relation (0.004 mag), and the uncertainties on the zero points of the V and K band photometries (0.010 mag and 0.015 mag, respectively). Combining these in quadrature, we obtain a total systematic error of 0.046 mag.



\section{Summary and conclusions}
We have obtained stellar parameters for the eclipsing binary OGLE-LMC-ECL-25658. Although this is an extragalactic object the absolute dimensions of the system, and the physical parameters of the component stars are determined with very high precision (better than 1\%). The system is composed of two giants of equal masses and similar temperatures but having significantly different radii and rates of axial rotation. As such this is probably an interesting case of differential stellar evolution of two identical stars in which some secondary evolution parameters (like e.g. initial rate of rotation) has led to the observed present day differences. We leave this problem for a future study, including a detailed comparison with evolutionary models. 

The distance to the system is measured to be $49.03\pm1.41$ kpc (total uncertainty), i.e. it has a fractional accuracy better than 3\%. The eclipsing binary is located relatively far from the barycenter of the LMC and lies on eastern side of the galaxy. Its position is exactly opposite to the systems analysed by P13 which are located close to the barycenter, and the western part of the LMC. Because of this our distance determination serves as a perfect check of the consistency of our method and the assumed spatial orientation of the LMC disc (van Marel's model). The distance to the LMC barycenter resulting from the application of the geometrical correction corresponding to position of the system is fully consistent with our previous result (P13, $d_{\rm LMC}=49.97\pm1.12$ kpc). With future $\sim10$ new eclipsing binaries to be analysed by our team we can provide very precise individual distance determinations to a total number of about 20 systems. Such a large sample
of systems, in tandem with an improved surface brightness-color relation our group is currently working on, will allow a significant improvement in the accuracy of the distance  to the LMC, and thus of the zero point of the cosmic distance scale.


\acknowledgments
We [S.E., D.G., W.G., G.P., M.G.] gratefully acknowledge financial support for this work from the BASAL Centro de Astrofisica y Tecnologias Afines (CATA) PFB-06/2007, and from the Millenium Institute of Astrophysics (MAS) of the Iniciativa Cientifica Milenio del Ministerio de Economia, Fomento y Turismo de Chile, project IC120009. Support from the Polish National Science Center grants MAESTRO
DEC-2012/06/A/ST9/00269 and OPUS DEC-2013/09/B/ST9/01551 is also acknowledged. The OGLE project has received funding from the National Science Centre,
Poland, grant MAESTRO 2014/14/A/ST9/00121 to AU. Grant NCN (Polish National Science Center) DEC-2011/03/B/ST9/02573 is acknowledge by IS.

\facility{Magellan:Clay (MIKE)}, \facility{ESO:3.6m (HARPS)}, \facility{NTT (SOFI)}

\end{document}